\documentclass[aps,prc,twocolumn,superscriptaddress,longbibliography]{revtex4-2}
\usepackage{graphicx}
\usepackage{amsmath}
\usepackage{amssymb}
\usepackage{bm}
\usepackage{xcolor}
\usepackage{hyperref}
\usepackage{booktabs}
\usepackage{multirow}
\usepackage{siunitx}
\usepackage{physics}
\newcommand{\Esum}{E_{\mathrm{sum}}}
\newcommand{\Easym}{E_{\mathrm{asym}}}
\newcommand{\thetaopen}{\theta_{\mathrm{open}}}
\newcommand{\geant}{Geant4}
\makeatletter
\newcommand{\AffLabel}[1]{%
  \ifcase#1\relax
  \or 1a% 1: INFN Pisa
  \or 2a% 2: INFN Pavia
  \or 2b% 3: University of Pavia
  \or 3a% 4: INFN Roma
  \or 3b% 5: University of Rome
  \or 1b% 6: University of Pisa
  \or 4%  7: PSI
  \fi
}%

\renewcommand\@affil@script[4]{%
 \@ifnum{#1=\z@}{}{%
  \par
  \begingroup
   \frontmatter@affiliationfont
   \textsuperscript{\AffLabel{#1}}%
   \ignorespaces#3%
   \@if@empty{#4}{}{\frontmatter@footnote{#4}}%
   \par
  \endgroup
 }%
}%

\def\@affil@present@script{%
 \let\@tempa\@empty
 \expandafter\@affil@present@script@\@affilID@temp\relax
}%
\def\@affil@present@script@#1{%
 \@ifx{\relax#1}{%
  \@ifx{\@tempa\@empty}{%
   \aftergroup\false@sw
  }{%
   \textsuperscript{\expandafter\@affilcommalabel\@tempa\relax\relax}%
   \aftergroup\true@sw
  }%
 }{%
  \@ifnum{#1=\z@}{}{\appdef\@tempa{{#1}}}%
  \@affil@present@script@
 }%
}%
\def\@affilcommalabel#1#2{%
 \@ifx{\z@#1}{%
  \@ifx{\relax#2}{}{\@affilcommalabel{#2}}%
 }{%
  \AffLabel{#1}%
  \@ifx{\relax#2}{}{%
   \@ifx{\z@#2}{\@affilcommalabel}{,\,\@affilcommalabel{#2}}%
  }%
 }%
}%
\makeatother

\begin{document}
\title{On the importance of cosmic-ray background in the Atomki anomaly}

% ---- Author / Affiliation block ----
% HB first (corresponding author -- marked automatically via \email),
% remaining authors alphabetical by surname.

\author{H.~Benmansour}
\email{hicham.benmans@gmail.com}
\affiliation{INFN, Sezione di Pisa, Largo Bruno Pontecorvo 3, I-56127 Pisa, Italy}

\author{G.~Boca}
\affiliation{INFN, Sezione di Pavia, Via Bassi 6, I-27100 Pavia, Italy}
\affiliation{University of Pavia, Department of Physics, Via Bassi 6, I-27100 Pavia, Italy}

\author{G.~Cavoto}
\affiliation{INFN, Sezione di Roma, Piazzale A.~Moro 2, I-00185 Rome, Italy}
\affiliation{University of Rome La Sapienza, Department of Physics, Piazzale A.~Moro 2, I-00185 Rome, Italy}

\author{M.~Chiappini}
\affiliation{INFN, Sezione di Pisa, Largo Bruno Pontecorvo 3, I-56127 Pisa, Italy}

\author{E.~G.~Grandoni}
\affiliation{INFN, Sezione di Pisa, Largo Bruno Pontecorvo 3, I-56127 Pisa, Italy}
\affiliation{University of Pisa, Department of Physics, Largo Bruno Pontecorvo 3, I-56127 Pisa, Italy}

\author{L.~Galli}
\affiliation{INFN, Sezione di Pisa, Largo Bruno Pontecorvo 3, I-56127 Pisa, Italy}

\author{G.~Gallucci}
\affiliation{INFN, Sezione di Pisa, Largo Bruno Pontecorvo 3, I-56127 Pisa, Italy}

\author{A.~Papa}
\affiliation{INFN, Sezione di Pisa, Largo Bruno Pontecorvo 3, I-56127 Pisa, Italy}
\affiliation{University of Pisa, Department of Physics, Largo Bruno Pontecorvo 3, I-56127 Pisa, Italy}
\affiliation{Paul Scherrer Institut PSI, CH-5232 Villigen, Switzerland}

\author{F.~Renga}
\affiliation{INFN, Sezione di Roma, Piazzale A.~Moro 2, I-00185 Rome, Italy}

\author{A.~Venturini}
\affiliation{INFN, Sezione di Pisa, Largo Bruno Pontecorvo 3, I-56127 Pisa, Italy}

\author{C.~Voena}
\affiliation{INFN, Sezione di Roma, Piazzale A.~Moro 2, I-00185 Rome, Italy}
\affiliation{University of Rome La Sapienza, Department of Physics, Piazzale A.~Moro 2, I-00185 Rome, Italy}

\date{\today}

% ─────────────────────────────────────────────────────────────────────────────
\begin{abstract}
We report \geant{}-based simulations of cosmic-ray muon backgrounds in models reproducing the geometries of the five-arm and six-arm $e^+e^-$ pair spectrometers operated by the Atomki group at Debrecen. Full detector geometries are
implemented for both setups, including position-sensitive
detectors and plastic scintillators. In both
configurations, cosmic muons generating two-arm coincidences produce
statistically significant excesses in the opening-angle distribution
whose positions are determined by the discrete azimuthal
arrangement of detector arms and the plastic scintillators dimensions: in the six-arm spectrometer a peak
appears near $140^\circ$ when the scintillator energy sum is selected
in the $^8$Be transition window ($\Esum \in [16, 20]$~MeV) and is
suppressed for energy-asymmetric pairs, while a distinct peak near
$120^\circ$ emerges in the $^4$He window ($\Esum \in [18, 22]$~MeV),
with no comparable enhancement in the respective background energy
region. These angles, asymmetry dependences, and background
characteristics show behaviours similar to those reported by the Atomki group as evidence for a 17~MeV boson. For the five-arm setup, a $140^\circ$  excess is also reproduced at $^8$Be energies and normalizing
internal pair conversion events and cosmic coincidences to typical
Atomki running conditions ($I_p = 1.0~\mu$A, $\sim$300~hours) yields
a cosmic-to-IPC ratio above unity in the signal window,
indicating that the cosmic background is a leading contribution to the event rate at the energies and angles of interest. While they do not settle the question of the origin of the Atomki excesses, these results highlight the critical importance of a robust cosmic-ray treatment in this type of measurements, and call for a detailed description of the beam-off studies supporting the search for anomalies in the $(\thetaopen, \Esum)$ distribution around the signal region.
\end{abstract}
\maketitle

% ─────────────────────────────────────────────────────────────────────────────
\section{\label{sec:intro}Introduction}

An anomalous excess in the $e^+e^-$ opening-angle distribution $\thetaopen$ from the
$^7$Li$(p,e^+e^-)^8$Be reaction was reported by Krasznahork{\'a}y
et al.\ in 2016~\cite{Krasznahorkay2016}, using a five-arm pair
spectrometer composed of multi-wire proportional chambers (MWPCs) and plastic scintillators, at the Atomki institute in Debrecen, Hungary. The excess appears as a bump above the expected Internal Pair Conversion (IPC) background~\cite{Rose1949,Rose1951,Rose1966}, near $\thetaopen\simeq 140^\circ$ for symmetric
pairs ($|E_{\mathrm{asym}}| < 0.5$) in the energy range corresponding
to the 18.15~MeV transition of $^8$Be. The signal was interpreted as
evidence for a new boson of mass $m_X \approx 17$~MeV (``X17'')
decaying to $e^+e^-$. In 2018, the group re-investigated the anomaly in $^8$Be, replacing the MWPCs with double-sided silicon strip detectors (DSSDs) and adding one additional arm. Again, a similar excess was observed~\cite{Krasznahorkay2018}. A subsequent measurement using the six-arm
spectrometer reported a kinematically compatible excess in the $^4$He system via
$^3$H$(p,e^+e^-)^4$He~\cite{Krasznahorkay2021}, with a peak near
$\thetaopen\simeq 115^\circ$.

The claim has generated considerable theoretical
interest~\cite{Feng2016,Ellwanger2016,Feng2017,Feng2020,Koch2021,Aleksejevs2021,Viviani2022}, but
independent experimental confirmation has remained elusive, the NA48/2~\cite{Batley2015} and NA64~\cite{Banerjee2020} experiments setting strong constraints on potential X17 couplings. In Ref.~\cite{Abraamyan2023}, Abraamyan et al. investigated the invariant mass spectra of photon pairs produced in dC, pC, and dCu collisions at various beam momenta. They reported a significant excess of events at an invariant mass of approximately 17~MeV/$c^2$, potentially compatible with the X17 hypothesis. The PADME
experiment at Frascati~\cite{Padme2025} annihilated a positron beam on a fixed target, aiming at producing X17 resonantly. The data were found to be consistent with the expected background in most of the investigated energy range. A deviation from background of approximately 2 standard deviations was seen for $\sqrt{s}\approx 16.90$ MeV. With a two-arm spectrometer inspired by Atomki, Anh et al.~\cite{Anh2024} attempted to reproduce the Atomki measurement using a Li$_2$O target irradiated with a proton beam. While no significant excess was observed at resonance ($E_p = 1030$~keV), they reported an excess over the expected IPC background at $E_p = 1225$~keV, corresponding to an opening angle of approximately $140^\circ$. The MEG II collaboration~\cite{Afanaciev2024} also investigated the $^7$Li$(p,e^+e^-)^8$Be reaction using a detector setup fundamentally different from that of Atomki, based on a magnetic spectrometer~\cite{CDCHpaper}. No excess above the expected IPC background was observed, and the results disfavor the Atomki X17 hypothesis with a significance of 1.5$\sigma$~\cite{Afanaciev2025}.

Since cosmic rays can create IPC-like coincidences and none of the cited Atomki experiments employed a fully efficient cosmic-ray veto, the analyses relied on cosmic-background subtraction using beam-off data recorded before and after the beam-on data-taking periods. The present work aims to quantitatively characterize the cosmic muon background in both Atomki geometries. We draw on the existing literature to simulate, as accurately as possible, both the five-arm (Krasznahork{\'a}y et al. 2016~\cite{Krasznahorkay2016} and Gulyás et al., 2016~\cite{Gulyas2016}) and six-arm (Krasznahork{\'a}y et al., 2021~\cite{Krasznahorkay2021}) spectrometers
in \geant{} 11.1~\cite{Geant4_2003,Geant4_2006,Geant4_2016}, propagating cosmic muons sampled from the Gaisser parametrization~\cite{Gaisser} through the full detector geometry. For the five-arm setup we additionally
simulate IPC events using the Zhang--Miller
model~\cite{Zhang2017} and normalize both components to the Atomki
beam and target conditions, in order to quantify the relative contribution
of each background source in the signal window.

The paper is organized as follows. Section~\ref{sec:setup} describes
the two spectrometer configurations and their implementation in the
simulation. Section~\ref{sec:simulation} details the physics modelling.
Sections~\ref{sec:6arm} and~\ref{sec:5arm} present the results for the
six- and five-arm setups respectively, the latter including the
IPC normalization study. Section~\ref{sec:discussion} discusses the findings and the limitations in interpreting the Atomki experiments. Section~\ref{sec:conclusions} summarizes the main points.

% ─────────────────────────────────────────────────────────────────────────────
\section{\label{sec:setup}The Atomki Pair Spectrometers}

\subsection{\label{sec:5arm_setup}2016 Five-Arm Spectrometer}

\begin{figure}[h]
  \centering
\includegraphics[width=0.98\columnwidth]{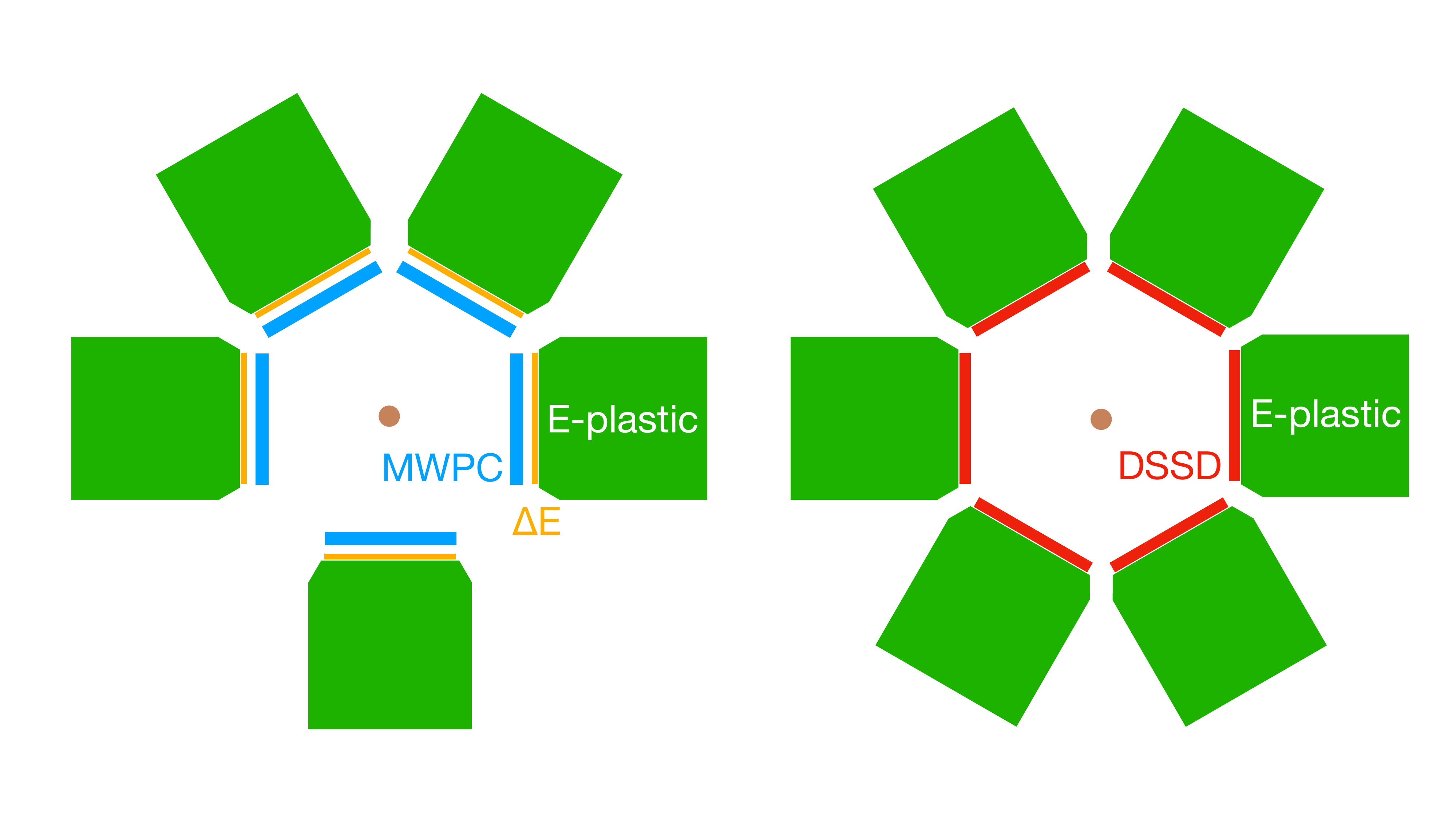}
  \caption{\label{fig:6arm_geom}
    (Left) Scheme of Atomki's five-arm spectrometer. Transverse
    view showing the five MWPC (blue) + $\Delta$E (orange) + E-plastic (green) telescopes at
    $\phi = 0^\circ, 60^\circ, 120^\circ, 180^\circ, 270^\circ$ around the target (brown). (Right) Scheme of Atomki's six-arm spectrometer. Transverse
    (X-Y) view showing the six DSSD (red) + E-plastic (green) telescopes at $60^\circ$ intervals around the target (brown). }
\end{figure}

The five-arm spectrometer employed in the original 
$^8$Be measurement is described in Refs.~\cite{Krasznahorkay2016,Gulyas2016}. A schematic view of the detector geometry is shown in Fig.~\ref{fig:6arm_geom}~(left).

Five identical telescopes are arranged at azimuthal angles
$\phi = 0^\circ, 60^\circ, 120^\circ, 180^\circ, 270^\circ$;
the $90^\circ$ gap between the last two arms breaks the otherwise
regular $60^\circ$ pattern and improves acceptance uniformity near
$90^\circ$.
Each telescope consists, from the beam axis outward, of a multi-wire
proportional chamber (MWPC), a thin $\Delta E$ scintillator, a thick
$E$ scintillator, and a passive light guide coupled to a
photomultiplier tube.

Three sensitive volumes are simulated per telescope.
Each MWPC consists of a $38 \times 45 \times 7$~mm$^3$ active gas
volume (Ar(80\%)+CO$_2$(20\%) at atmospheric pressure) flanked by two
0.1~mm cathode boards (PCB, modelled as Bakelite), with its inner face at $r = 35$~mm; the
gas depth and composition are stated in Ref.~\cite{Gulyas2016}, while
the active face of $38 \times 45$~mm$^2$ is inferred from the
$\Delta E$ dimensions since no explicit MWPC face size is given in
either reference.
The dimensions of the $\Delta E$ detectors ($38 \times 45 \times 1$~mm$^3$, EJ-200 equivalent) and the $E$ calorimeters ($78 \times 60 \times 70$~mm$^3$, EJ-200 equivalent) were taken from Ref.~\cite{Krasznahorkay2016}. Their radial positions were chosen to maximize the geometry's compactness.
To avoid the geometric overlap that a single rectangular $E$ block
would produce between $60^\circ$-adjacent arms at small radii, the
$E$ volume is implemented as a trapezoidal front section joined to a
rectangular body, consistent with the bevelled shape visible in
Fig.~5 of Ref.~\cite{Gulyas2016}.

%Among the passive elements, the Lucite light guide behind each $E$ detector is included. The light-guide geometry is not described in either reference and is assigned a conservative placeholder.

The following passive elements are not simulated: the
carbon-fibre beam pipe wall, the MWPC plastic housing walls and the target assembly. The original $^8$Be measurement used a LiO$_2$ target.

\subsection{\label{sec:6arm_setup}2021 Six-Arm Spectrometer}
The six-arm spectrometer, used for both an updated $^8$Be
measurement~\cite{Krasznahorkay2018} and the $^4$He
result~\cite{Krasznahorkay2021}, replaces the MWPCs with double-sided
silicon strip detectors (DSSDs). A schematic view of the detector geometry is shown in Fig.~\ref{fig:6arm_geom}~(right).
Six identical telescopes are arranged at
$\phi = n \times 60^\circ$, $n = 0,\ldots,5$.
Each telescope consists, from the beam axis outward, of a DSSD,
a plastic scintillator, and a passive light guide coupled to a
photomultiplier tube.

Two sensitive volumes are simulated per telescope. The DSSD ($50 \times 50$~mm$^2$, 500~$\mu$m thick, silicon) sits in
front of the plastic scintillator ($82 \times 86 \times 80$~mm$^3$,
EJ-200 equivalent).
The detector dimensions are taken from Ref.~\cite{Krasznahorkay2021};
the radial position of the assembly was set so that the scintillator blocks of adjacent arms do not physically overlap, given the $60^\circ$ angular
separation.

The beam-pipe vacuum is
included, but the carbon-fibre wall is not simulated; the region
between the vacuum and the DSSD inner face is filled with air.
The target assembly and all other passive structural elements are
likewise not simulated. The $^4$He data were taken with a $^3$H target.

Tables~\ref{tab:geom6} and~\ref{tab:geom5} collect the geometric
parameters implemented in the two simulations, taken or inferred from Refs.~\cite{Krasznahorkay2016,Gulyas2016,Krasznahorkay2021} or chosen when unavailable.

\begin{table}[h]
\caption{\label{tab:geom6}Six-arm spectrometer geometry.}
\begin{ruledtabular}
\begin{tabular}{lll}
Component     & Parameter          & Value \\
\hline
%Beam pipe     & Inner radius       & 26~mm \\
%              & Wall               & CF epoxy, 1~mm \\
%              & Half-length        & 200~mm \\
\hline
DSSD          & Active area        & $50\times 50$~mm$^2$ \\
              & Thickness          & 500~$\mu$m Si \\
              & Threshold          & 50~keV \\
\hline
Scintillator  & Dimensions         & $82\times 86\times 80$~mm$^3$ \\
              & Material           & Vinyltoluene \\
              & Threshold          & 100~keV \\
\hline
%Light guide   & Dimensions         & $60\times 40\times 40$~mm$^3$ \\
%              & Material           & Lucite \\
\hline
Arms          & Number             & 6 \\
              & $\phi$             & $n\times 60^\circ$, $n=0\ldots 5$ \\
\end{tabular}
\end{ruledtabular}
\end{table}

\begin{table}[h]
\caption{\label{tab:geom5}Five-arm spectrometer geometry.}
\begin{ruledtabular}
\begin{tabular}{lll}
Component     & Parameter          & Value \\
\hline
%Beam pipe     & Inner radius       & 34.2~mm \\
%              & Wall               & Vacuum (CF omitted) \\
\hline
MWPC          & Active area        & $38\times 45$~mm$^2$ \\
              & Gas                & Ar(80\%)+CO$_2$(20\%) \\
              & Gas gap            & 7~mm \\
              & Cathode boards     & 0.1~mm Bakelite \\
\hline
$\Delta E$    & Dimensions         & $38\times 45\times 1$~mm$^3$ \\
              & Material           & Vinyltoluene \\
              & Threshold          & 50~keV \\

\hline
$E$-plastic   & Dimensions         & $78\times 60\times 70$~mm$^3$ \\
              & Material           & Vinyltoluene \\
              & Threshold          & 100~keV \\
\hline
%Light guide   & Dimensions         & $40\times 40\times 40$~mm$^3$ \\
%              & Material           & Lucite \\
\hline
Arms          & Number             & 5 \\
              & $\phi$             & $0^\circ,60^\circ,120^\circ,180^\circ,270^\circ$ \\
\end{tabular}
\end{ruledtabular}
\end{table}
% ─────────────────────────────────────────────────────────────────────────────
\section{\label{sec:simulation}Simulation Framework}

\subsection{\label{sec:geant4}Geant4 Implementation}

Separate \geant{} 11.1 simulations were implemented for each spectrometer. We used the \texttt{FTFP\_BERT\_EMY} reference physics list (\texttt{FTFP\_BERT} for hadronic interactions combined with \texttt{G4EmStandardPhysics\_option3} for electromagnetic interactions). Energy deposits are accumulated step-by-step in \texttt{SteppingAction}, which identifies sensitive volumes by logical-volume name and telescope copy number. Hit thresholds are 1~keV for MWPCs, 50~keV for DSSDs, and 100~keV for plastic scintillators. Per-event quantities are written to ROOT~\cite{ROOT}
ntuples.

\subsection{\label{sec:cosmics_gen}Cosmic Muon Generator}

Muons are generated from a $2000 \times 2000$~mm$^2$ plane 600~mm above
the apparatus. The angular distribution follows a cosine law within
$0^\circ$--$70^\circ$ of the vertical. The energy spectrum is sampled
from a piecewise-linear approximation to the Gaisser
formula~\cite{Gaisser}:
\begin{equation}
\frac{d^2\Phi_\mu}{dE_\mu\,d\Omega}
=
0.14\,E_\mu^{-2.7}
\left(
\frac{1}{1+\dfrac{1.1E_\mu\cos\theta}{115}}
+
\frac{0.054}{1+\dfrac{1.1E_\mu\cos\theta}{850}}
\right),
\label{eq:gaisser}
\end{equation}
(in units of cm$^{-2}$ s$^{-1}$ sr$^{-1}$ GeV$^{-1}$), between
100~MeV and 100~GeV. The integrated vertical flux
$\Phi_\mu \approx 1$~cm$^{-2}$~min$^{-1}$ gives a rate of
$R_\mu \approx 700$~Hz through the source plane.

\subsection{\label{sec:observables}Reconstructed Observables}

The analysis uses four main per-event quantities:
\begin{description}
  \item[$N_{\mathrm{hit}}$] number of arms being hit. An arm is considered hit when all active elements (DSSD+plastic or MWPC+$\Delta$E+plastic) record an energy deposit above the corresponding thresholds;
  \item[$\Esum$] scalar sum of energies deposited in the two arms for events with $N_{\mathrm{hit}} = 2$;
  \item[$\Easym$] energy asymmetry (considering $E_1$ and $E_2$ the energies deposited coincidentally in the arms when $N_{\mathrm{hit}} = 2$),
    \begin{equation}
      \Easym = \frac{E_1 - E_2}{E_1 + E_2}, \quad E_1 \geq E_2;
      \label{eq:Easym}
    \end{equation}
  \item[$\thetaopen$] opening angle between the energy-weighted
        centroids of the two hit DSSDs (or MWPCs) as seen from the target centre. For the $^4$He measurement, the 25~mm-offset between the target and the apparatus centre is accounted for.
    \begin{equation}
      \thetaopen = \arccos\!\bigl(\hat{r}_1 \cdot \hat{r}_2\bigr).
      \label{eq:theta}
    \end{equation}
\end{description}
These definitions match those used in the Atomki
analyses~\cite{Krasznahorkay2016,Krasznahorkay2021}.

\subsection{\label{sec:ipc_gen}IPC Generator}

IPC events for the five-arm normalization study
(Sec.~\ref{sec:normalization}) are generated with the full Zhang--Miller
model~\cite{Zhang2017} for $^7$Li$(p,e^+e^-)^8$Be at the 1030~keV
resonance, comprising E1 direct capture, M1 resonances, and E2
contributions with Coulomb phase factors. The differential cross section
$d^2\sigma / dE_{e^+}\, d(\cos\theta_{e^+e^-})$ is sampled via a
$50 \times 50$ acceptance-rejection table. The transverse vertex
position is smeared with a 2D Gaussian of $\sigma = 2$~mm.

% ─────────────────────────────────────────────────────────────────────────────
\section{\label{sec:6arm}Six-Arm Spectrometer}

All the results in this section derive from a cosmic-only simulation. 
\subsection{\label{sec:6arm_mult}Hit Multiplicity and Energy Deposits}

The simulated geometry is sketched in
Fig.~\ref{fig:6arm_geom}~(right). Approximately 0.2‰ of generated cosmic muons produce a two-DSSD coincidence.

\begin{figure}[h]
  \centering
  \includegraphics[width=\columnwidth]{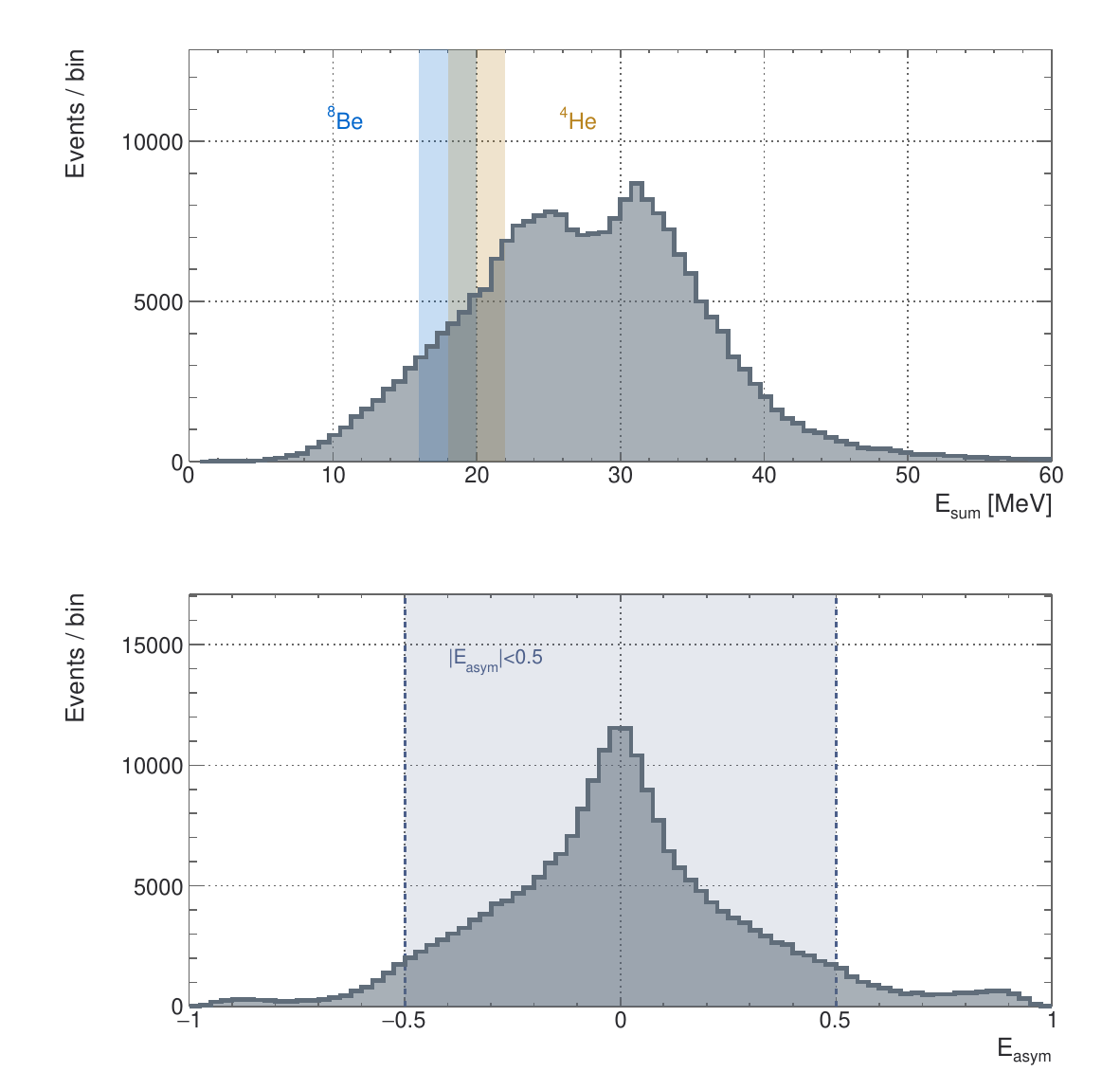}
  \caption{\label{fig:esum_easym}
    (Top) $\Esum$ distribution (sum of both $\Delta$E+E energy deposits) for
    cosmic events with $N_{\mathrm{hit}} = 2$ in the six-arm
    simulation. The shaded bands indicate the $^8$Be and $^4$He signal windows used in Refs.~\cite{Krasznahorkay2016,Krasznahorkay2021}.
    (Bottom) $\Easym$ distribution for the same sample; the dashed lines
    mark $|\Easym| = 0.5$.}
\end{figure}

In two-hit events the DSSD energy deposits are a few hundreds~keV, consistent
with minimum-ionizing-particle loss in 500~$\mu$m of Si, while the
scintillator deposits up to 30~MeV per arm. The $\Esum$
distribution and $\Easym$ distribution resulting from our cosmic muons generation are shown in
Fig.~\ref{fig:esum_easym}. The $\Esum$ distribution peaks close to 30~MeV with a left tail down to 10~MeV, showing cosmic muons can mimic $^8$Be and $^4$He IPC energy deposits.  The majority of cosmic coincidences also satisfy
$|\Easym| < 0.5$, similarly to Atomki's observed excesses.

\subsection{\label{sec:6arm_pairs}Arm-Pair Geometry and Opening Angles}

\begin{figure}[h]
  \centering
  \includegraphics[width=\columnwidth]{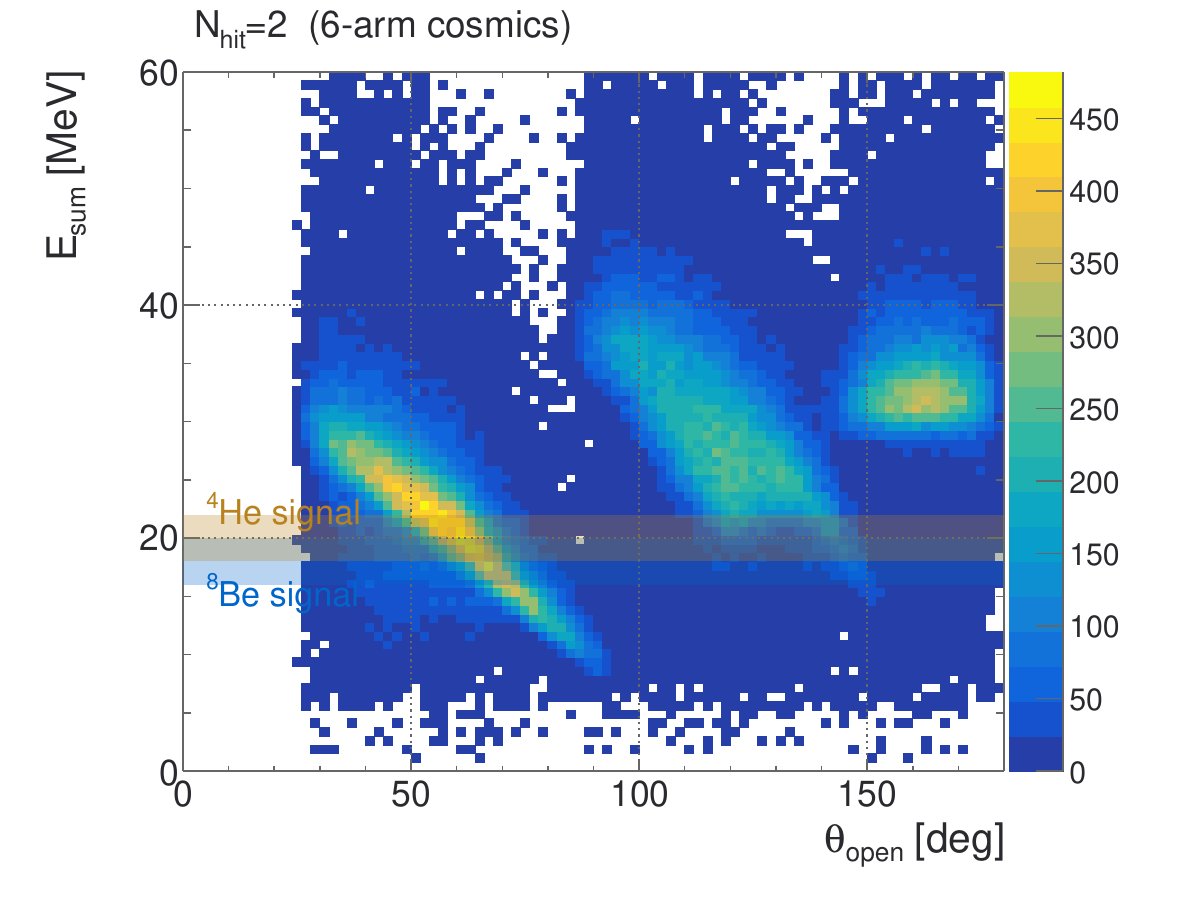}
  \caption{\label{fig:esum_angle_2d_6arm}
    Two-dimensional $(\thetaopen, \Esum)$ distribution for cosmic
    two-arm events in the six-arm simulation. The $^8$Be signal band
    ($\Esum \in [16, 20]$~MeV, blue) and $^4$He signal band
    ($\Esum \in [18, 22]$~MeV, orange) are indicated.}
\end{figure}

With six arms at $60^\circ$ spacing, three distinct arm-pair types
arise in two-arm cosmic coincidences:
\begin{itemize}
  \item \emph{Opposite} ($\Delta\phi = 180^\circ$): 20\% of events,
        $\thetaopen \simeq 160^\circ$--$180^\circ$;
  \item \emph{Adjacent} ($\Delta\phi = 60^\circ$): 41\%,
        $\thetaopen \simeq 60^\circ$--$80^\circ$;
  \item \emph{Next-to-adjacent} ($\Delta\phi = 120^\circ$): 38\%,
        $\thetaopen \simeq 110^\circ$--$140^\circ$.
\end{itemize}
The adjacent and next-to-adjacent two-hit configurations were found to be dominant. The $\Esum$ vs $\thetaopen$ distribution is shown in Fig.~\ref{fig:esum_angle_2d_6arm}. Three main populations are seen, each corresponding to one of the two-hit configurations. The distribution reveals a negative correlation between the two observables for cosmic two-arm coincidences involving adjacent arms ($\Delta\phi = 60^\circ$) and next-to-adjacent arms ($\Delta\phi = 120^\circ$). This correlation arises from the interplay between track geometry and hit-position reconstruction. Consider one cosmic ray hitting two (and only two) next-to-adjacent arms. A muon traversing both scintillators centrally deposits energy over a long path length (high $\Esum$) but produces hit centroids on the two DSSDs that are displaced, reducing the reconstructed opening angle below the nominal $120^\circ$ azimuthal separation. Conversely, a muon clipping the outer edges of the same arm pair — passing closer to the beam axis — deposits less energy (low 
$\Esum$) and produces increased $\thetaopen$, above $120^\circ$. The resulting negative $\Esum$--$\thetaopen$ correlation means that the Atomki signal selection window, defined at $\Esum$ values of [16,20]~MeV, preferentially selects a specific range of opening angles from cosmic background, which cannot be a priori distinguished from IPC events. The correlation between $\Esum$ and $\thetaopen$ should therefore be a key observable in the interpretation of the Atomki anomaly.

\subsection{\label{sec:6arm_8be}$^8$Be Signal Window}

\begin{figure}[h]
  \centering
  \includegraphics[width=\columnwidth]{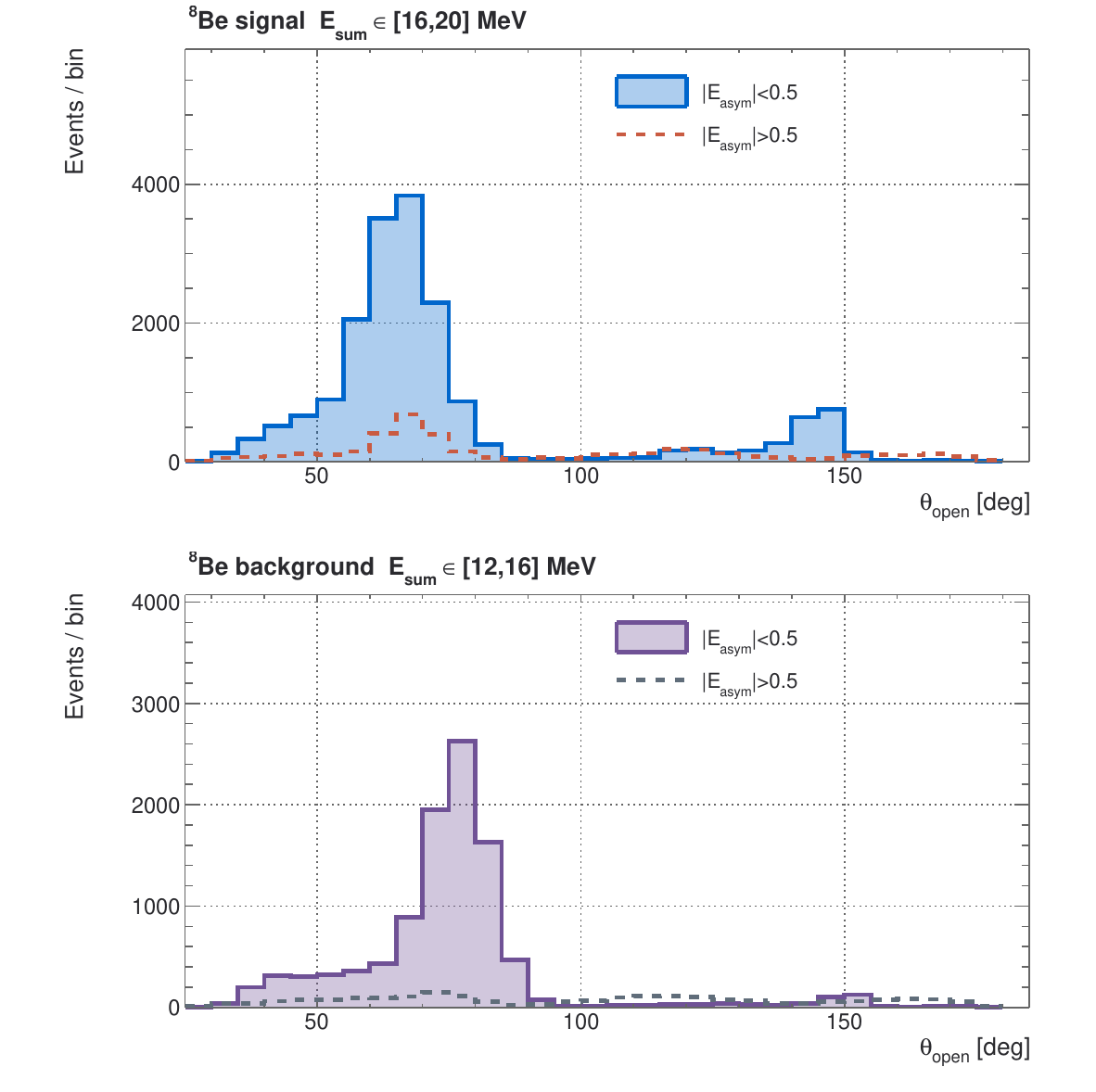}
  \caption{\label{fig:8be_proj_6arm}
    (Top) $\thetaopen$ distribution in the six-arm simulation for cosmic events
    with $\Esum \in [16,20]$~MeV. Symmetric pairs ($|\Easym| < 0.5$,
    solid blue filled) show a concentration near $140^\circ$; asymmetric pairs
    ($|\Easym| > 0.5$, red dashed) do not. (Bottom) $\thetaopen$ distribution in the six-arm simulation for cosmic events
    with $\Esum \in [12,16]$~MeV. Symmetric pairs ($|\Easym| < 0.5$)
    in solid purple filled. In this case, the excess near $140^\circ$ is suppressed by a factor 5. Asymmetric pairs
    ($|\Easym| > 0.5$) in grey dashed. The insert reproduces the
    Atomki analysis conditions from Ref.~\cite{Krasznahorkay2018} for comparison.}
\end{figure}

Selecting $\Esum \in [16, 20]$~MeV and $|\Easym| < 0.5$ (symmetric pairs), the cosmic $\thetaopen$ distribution shows a peak near $140^\circ$, as shown in Fig.~\ref{fig:8be_proj_6arm}~(top). The peak arises from the next-to-adjacent arm configuration: at these energies the two
scintillators share the muon energy roughly equally (hence symmetric $\Easym$), and the geometrical opening angle of a $\Delta\phi = 120^\circ$ pair is $\sim 140^\circ$. For asymmetric pairs ($|\Easym| > 0.5$, dashed red line) the peak vanishes; tracks producing strongly unequal energy sharing tend to connect arms at smaller azimuthal separations. The $\Esum \in [12, 16]$~MeV selection is shown in Fig.~\ref{fig:8be_proj_6arm}~(bottom). In this range, the excess is reduced by a factor of 5 with respect to the signal region.

The cosmic-ray simulation exhibits three of the same key features (peak position, reduced excess for asymmetric pairs and low-energy pairs) emphasized in Atomki's X17 analysis, indicating that cosmic coincidences can closely mimic the expected signal topology and must therefore be carefully characterized to distinguish them from any potential signal.

It should also be noted that these energy windows also select cosmic rays hitting adjacent arms, leading to opening angles near $\Delta\phi = 60^\circ$. However, in this low-angle region, IPC and EPC (photon External Pair Conversion) rates are much larger and cosmic events represent a smaller relative contribution than at $140^\circ$. 

\subsection{\label{sec:6arm_4he}$^4$He Signal Window}

\begin{figure}[h]
  \centering
 \includegraphics[width=\columnwidth]{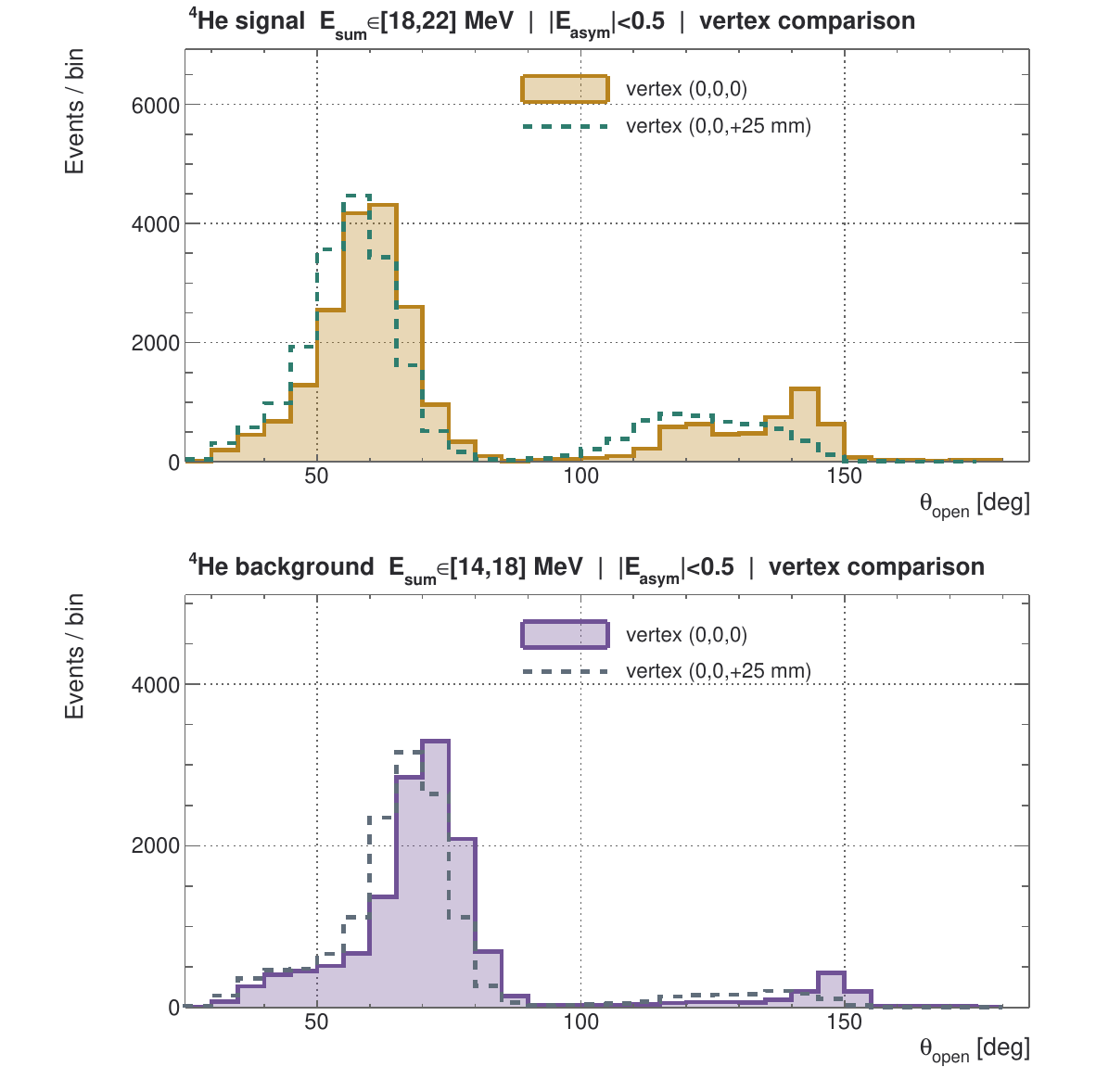}
  \caption{\label{fig:4he_proj_6arm}
    $\thetaopen$ distributions in the six-arm simulation for $|\Easym| < 0.5$ for the $^4$He
    signal window $\Esum \in [18,22]$~MeV (top) and the background window
    $\Esum \in [14,18]$~MeV (bottom). Opening angles seen from the apparatus center (0,0,0) in filled solid line and seen from the target center (0,0,+25~mm) in dashed line. The target offset in Atomki's $^4$He experiment leads to a cosmic excess peaking near 120$^\circ$ in the signal window. It is largely suppressed at lower $\Esum$. The insert reproduces the
    Atomki analysis conditions from Ref.~\cite{Krasznahorkay2021} for comparison.}
\end{figure}

In Ref.~\cite{Krasznahorkay2021}, the Atomki group defines the signal energy window as [18, 22]~MeV for the $^3$H$(p,e^+e^-)^4$He reaction, with transitions near 20~MeV. For $\Esum \in [18,22]$~MeV and $|\Easym|<0.5$, the $\thetaopen$ distribution shown in Fig.~\ref{fig:4he_proj_6arm} (top) exhibits two peaks above $100^\circ$: one near $140^\circ$, corresponding to the excess already observed at the $^8$Be transition energies, and another near $120^\circ$. At these slightly higher energies, an additional next-to-adjacent topology appears, visible in Fig.~\ref{fig:esum_angle_2d_6arm} around ($120^\circ$, 21~MeV), giving rise to the second peak. The excess around $120^\circ$ observed in our simulation shares a similar topology to the feature reported by Atomki. Unlike the $^8$Be analysis, the $\Easym$ observable was not used in the $^4$He study, and the interval $14~\mbox{MeV} < \Esum < 18$~MeV was instead taken as a background region. As shown in Fig.~\ref{fig:4he_proj_6arm} (bottom), this background window exhibits a strongly suppressed angular excess.

Because the $^3$H target was located inside a cooling pipe, it had to be positioned 25~mm downstream of the detector center. Reconstructing the opening angle from this displaced position shifts the $\theta_{\mathrm{open}}$ distribution towards smaller values, as illustrated by the dashed green curve in Fig.~\ref{fig:4he_proj_6arm}. This target offset therefore produces a broad enhancement in the $110^\circ$--$140^\circ$ region. In addition, the cooling pipe limited the maximum accessible opening angle in the $^4$He Atomki experiment to approximately $140^\circ$, and the $\theta_{\mathrm{open}}$ distributions reported in Ref.~\cite{Krasznahorkay2021} were presented only up to $135^\circ$. As a result, the higher-angle part of the structure predicted by our simulation lies outside the published range. Nevertheless, the last displayed bin at $130^\circ$ in Fig.~7 of Ref.~\cite{Krasznahorkay2021} still shows a residual excess above the expected background, consistent with the contribution from cosmic-ray coincidences predicted by our simulation.

% ─────────────────────────────────────────────────────────────────────────────
\section{\label{sec:5arm}Five-Arm Spectrometer}

Sections~\ref{sec:5arm_mult},~\ref{sec:5arm_esum_angle} and ~\ref{sec:5arm_8be} reproduce the cosmic-ray-only study previously performed for the six-arm spectrometer, now applied to the five-arm configuration. Sections~\ref{sec:ipc_acceptance},~\ref{sec:normalization} and ~\ref{sec:signal_region} then compare the relative contributions of IPC and cosmic-ray backgrounds to assess the actual impact of cosmic coincidences on the Atomki data acquisition and analysis.

\subsection{\label{sec:5arm_mult}Hit Multiplicity and Energy Deposits}

We now study two-arm cosmic coincidences in the five-arm setup. The simulated geometry is sketched in Figure~\ref{fig:6arm_geom}~(left).

The $\Esum$ distribution in two-arm events peaks
near 15-20~MeV with a tail extending to $\sim$5~MeV---wider than in the six-arm case, reflecting the smaller E-plastic size.
The $\Easym$ distribution is again sharply peaked at zero, with the majority
of events in the $|\Easym| < 0.5$ interval.

\subsection{\label{sec:5arm_esum_angle}Energy Sum vs.\ Opening Angle}

\begin{figure}[h]
  \centering
  \includegraphics[width=\columnwidth]{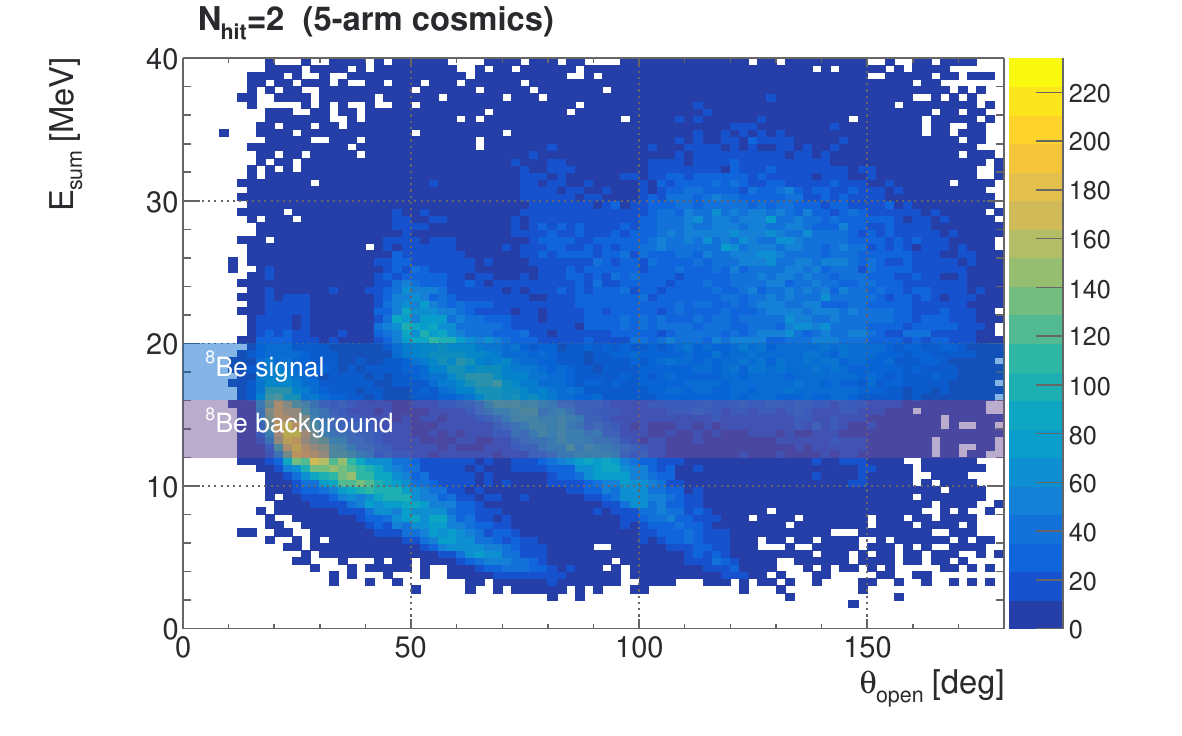}
  \caption{\label{fig:esum_angle_2d}
    Two-dimensional $(\thetaopen, \Esum)$ distribution for cosmic
    two-arm events in the five-arm simulation. The $^8$Be signal band
    ($\Esum \in [16, 20]$~MeV, blue) and background band
    ($\Esum \in [12, 16]$~MeV, purple) are indicated.
    At $\Esum = 18$~MeV the cosmic distribution peaks near
    $\thetaopen \simeq 140^\circ$.}
\end{figure}

Figure~\ref{fig:esum_angle_2d} shows the two-dimensional $(\thetaopen, \Esum)$ distribution for two-arm cosmic events in the five-arm simulation. A strong negative correlation is visible: lower $\Esum$ events tend to have larger opening angles. Again, this follows from
geometry, as previously explained. Note that in this case four two-hit configurations are accessible to the cosmic rays, the adjacent double-hit for telescopes separated by $\Delta\phi=60^\circ$ ($\thetaopen \simeq 20^\circ$--$70^\circ$), the adjacent one for telescopes separated by $\Delta\phi=90^\circ$ ($\thetaopen \simeq 50^\circ$--$100^\circ$), the next-to-adjacent one for cosmics hitting one of the two top telescopes and the bottom one $\Delta\phi=150^\circ$ ($\thetaopen \simeq 110^\circ$--$180^\circ$) and the next-to-adjacent one for telescopes separated by $\Delta\phi=120^\circ$ ($\thetaopen \simeq 90^\circ$--$130^\circ$).

\subsection{\label{sec:5arm_8be}$^8$Be Signal and Background Windows}

\begin{figure}[h]
  \centering
  \includegraphics[width=\columnwidth]{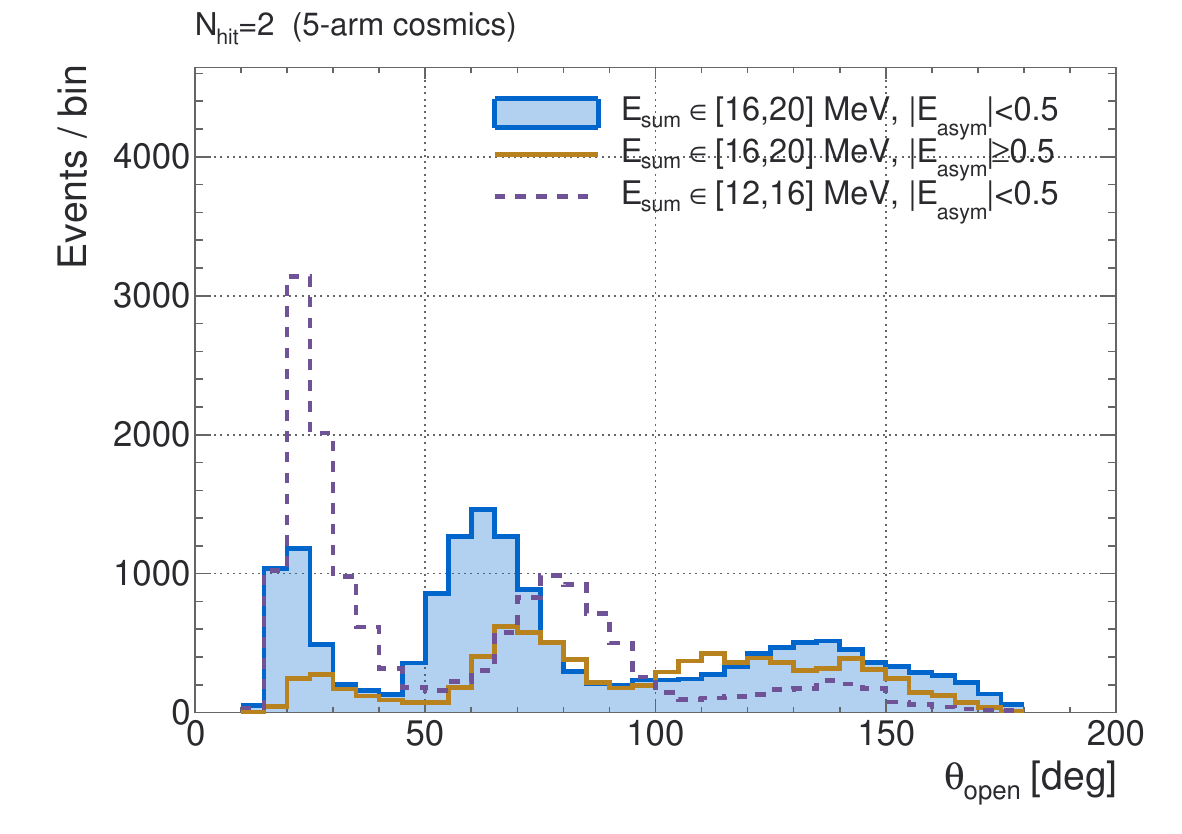}
  \caption{\label{fig:8be_5arm}
    $\thetaopen$ distributions in the five-arm cosmic simulation. Signal window $\Esum \in [16, 20]$~MeV for symmetric ($|\Easym| < 0.5$, filled solid blue) and asymmetric ($|\Easym| > 0.5$, solid orange) pairs. Background window
    $\Esum \in [12, 16]$~MeV ($|\Easym| < 0.5$, dashed purple).}
\end{figure}

Selecting $\Esum \in [16, 20]$~MeV and $|\Easym| < 0.5$ from the
five-arm cosmic simulation, the $\thetaopen$ distribution develops an
excess near $130^\circ$--$150^\circ$ (Fig.~\ref{fig:8be_5arm}), driven
by the $150^\circ$ next-to-adjacent arm topology. For asymmetric pairs the excess is slightly reduced but
still present, because the asymmetric five-arm layout (with the
$90^\circ$ gap) allows a somewhat wider variety of track topologies
than the regular six-arm case. In the background window
$\Esum \in [12, 16]$~MeV, the angular peak is reduced by a factor of 2.

\subsection{\label{sec:ipc_acceptance}IPC Simulation and Apparatus Acceptance}

\begin{figure}[h]
  \centering
  \includegraphics[width=\columnwidth]{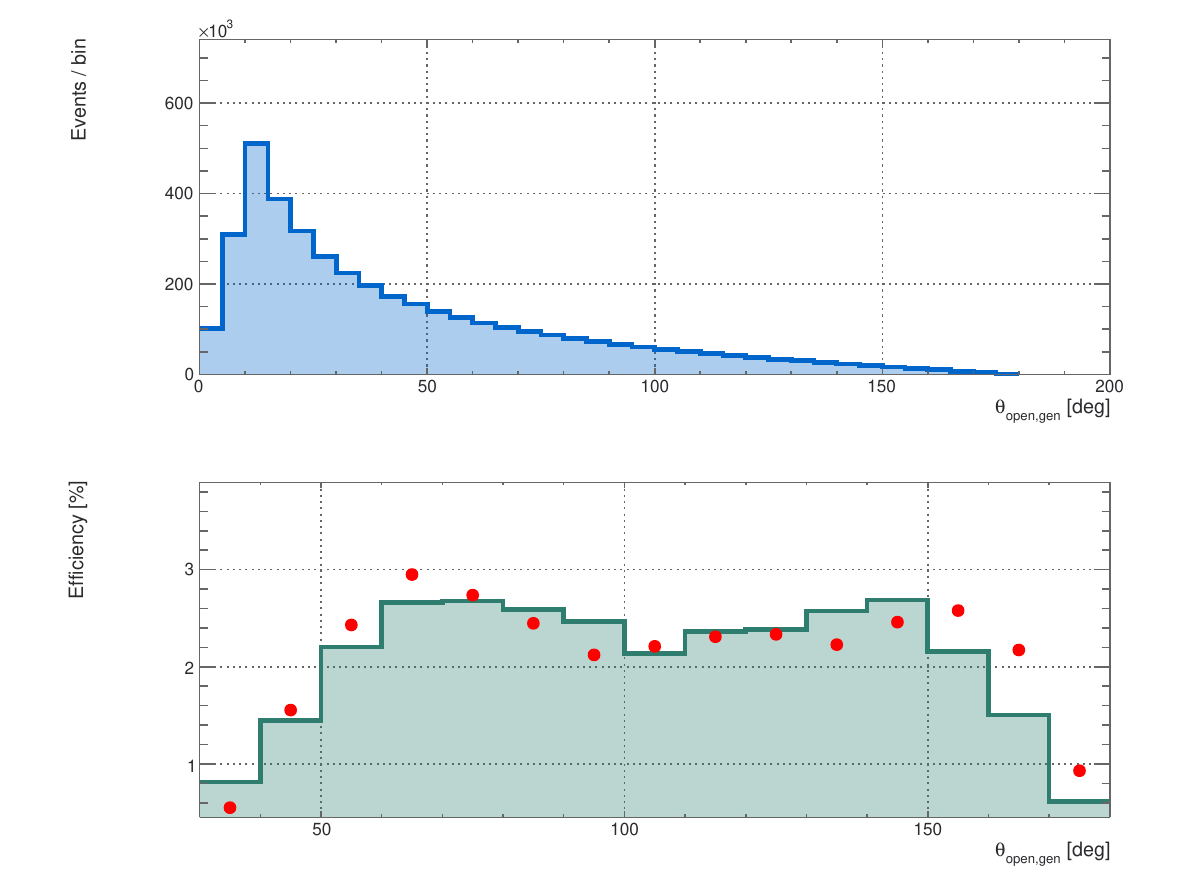}
  \caption{\label{fig:acceptance}
    Generated opening angle distribution (top) and apparatus acceptance (bottom) for IPC events in the five-arm simulation, including the $\sin\theta$ phase-space factor. The simulated acceptance (filled solid green) is compared to the measured response from Ref.~\cite{Gulyas2016}, Fig.~7 (filled red circles), normalized to the same area.}
\end{figure}

From $4\times10^6$ simulated $^8$Be IPC events at a transition energy of 18.15~MeV and a center-of-mass energy of $E_{\mathrm{CM}} = 0.895$~MeV (corresponding to a proton beam energy of 1.03~MeV on a $^7$Li target at rest),
$\approx 2.5\%$ produce a valid two-MWPC hit coincidence. The generated angular opening following Zhang-Miller model at this energy is shown in Fig.~\ref{fig:acceptance}~(top).
The apparatus acceptance, defined as the ratio of reconstructed to
generated opening-angle distributions (after including the $\sin\theta_{\mathrm{open,gen}}$ factor), shows two peaks at $60^\circ$ and $140^\circ$
(Fig.~\ref{fig:acceptance}~(bottom)). This is in qualitative agreement with the
efficiency curve measured by the Atomki group and shown in Fig.~7 from Ref.~\cite{Gulyas2016} (filled red circles in our figure), providing a validation of our simulated detector geometry. 

\subsection{\label{sec:normalization}Normalization and Combined Background}

\subsubsection{Reaction rates}

In Ref.~\cite{Krasznahorkay2016}, the Atomki group used a 5 MV Van de Graaff to send a proton beam with a current
of 1.0~$\mu$A onto a
700 $\mu$g/cm$^2$-thick LiO$_2$ target. It corresponds to a proton rate of
$\dot{N}_p = 6.24 \times 10^{12}$~p/s. The $^7$Li$(p,\gamma)^8$Be cross section at the 1030~keV resonance is approximately 20~$\mu$b. To account for both the proton energy loss in the target and the energy dependence of the cross section, the energy-dependent curve $\sigma(E)$ from Ref.~\cite{Gysbers2023} was used. For a
LiO$_2$ target of surface density $700~\mu$g/cm$^2$
($n = 1.0 \times 10^{19}$~atoms/cm$^2$, thickness $x = 3.5~\mu$m,
$|dE/dx| = 300$~MeV/cm) the reaction probability is
\begin{equation}
  P = \frac{n}{x\,|dE/dx|} \int \sigma(E)\,dE
    \approx 1.5 \times 10^{-10},
  \label{eq:P}
\end{equation}
giving a generated photon rate $R_\gamma = \dot{N}_p \cdot P \approx 1$~kHz and
an IPC rate $R_{\mathrm{IPC}} \approx 4$~Hz. %A cross-check with MEG~II in-beam data at $I_p = 10~\mu$A using their LiPON target predicts $\sim$20~Hz of 18~MeV photons in their BGO calorimeter placed 70~cm away from the target, consistent with their observation.

\subsubsection{Equivalent running time}

\begin{figure}[h]
  \centering
  \includegraphics[width=\columnwidth]{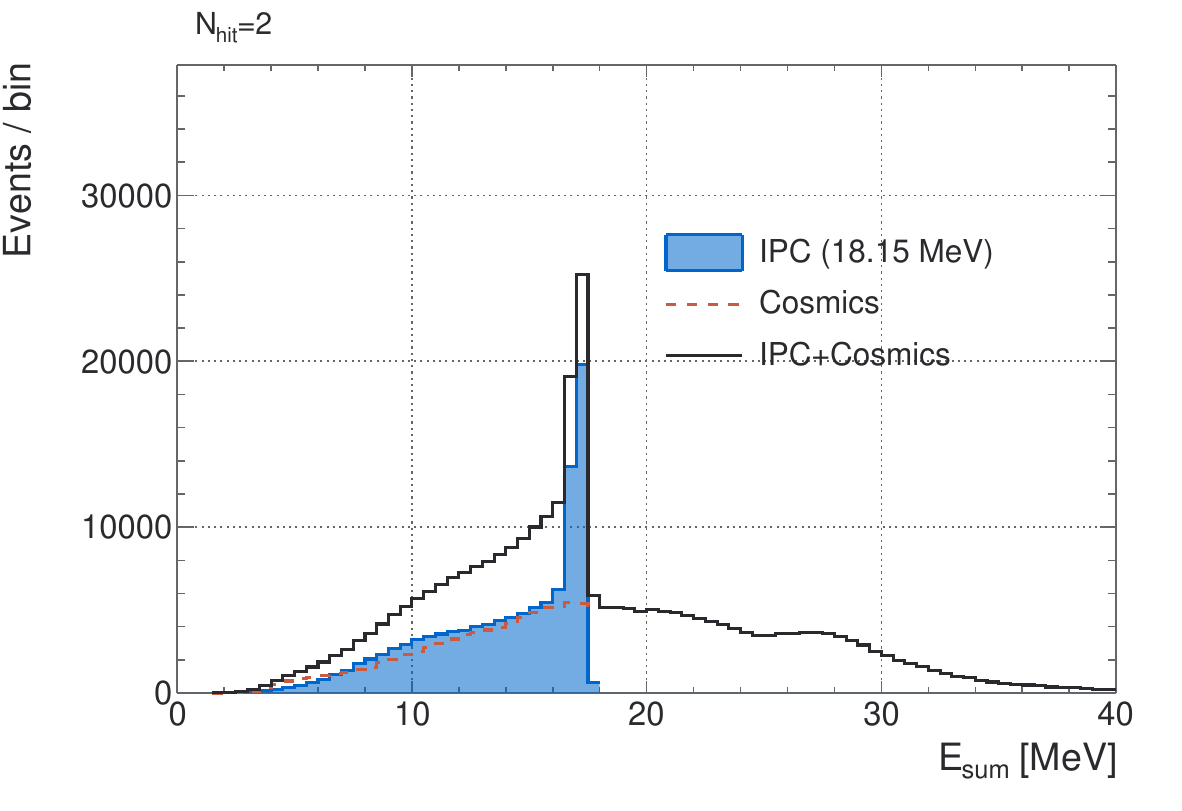}

  \caption{\label{fig:esum_combined}
    $\Esum$ spectra for IPC (solid blue filled) and cosmic (dashed red) events normalized
    to 300~hours of Atomki conditions ($I_p = 1~\mu$A,
    $R_\mu = 700$~Hz). The sum of both contributions is shown in solid back. The IPC peak at 18.15~MeV sits on a
    non-negligible cosmic continuum which represents approximately 50\% of the total number of events in the $[16, 20]$~MeV window.}
\end{figure}

Both cosmic rays and $^8$Be(18.1~MeV) IPC were then simulated to correspond to about 300~hours of Atomki's data-taking in Ref.~\cite{Krasznahorkay2016}'s conditions. The resulting $\Esum$ spectra are shown in Fig.~\ref{fig:esum_combined}.
The IPC component produces a narrow peak with a left tail corresponding to the E-plastics leakage; the cosmic
component contributes a broad continuum peaking near 15-20~MeV, with the $^8$Be signal window containing $\approx$50\% of cosmic events.

\subsection{\label{sec:signal_region}Signal Region}

\begin{figure}[h]
  \centering
  \includegraphics[width=\columnwidth]{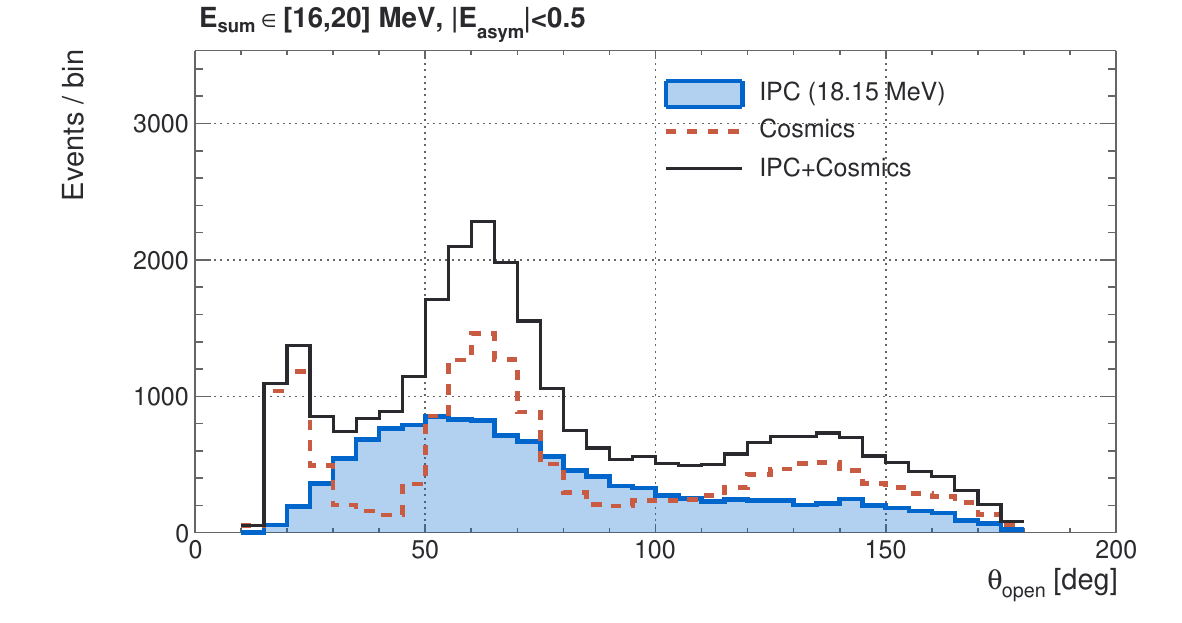}
\includegraphics[width=\columnwidth]{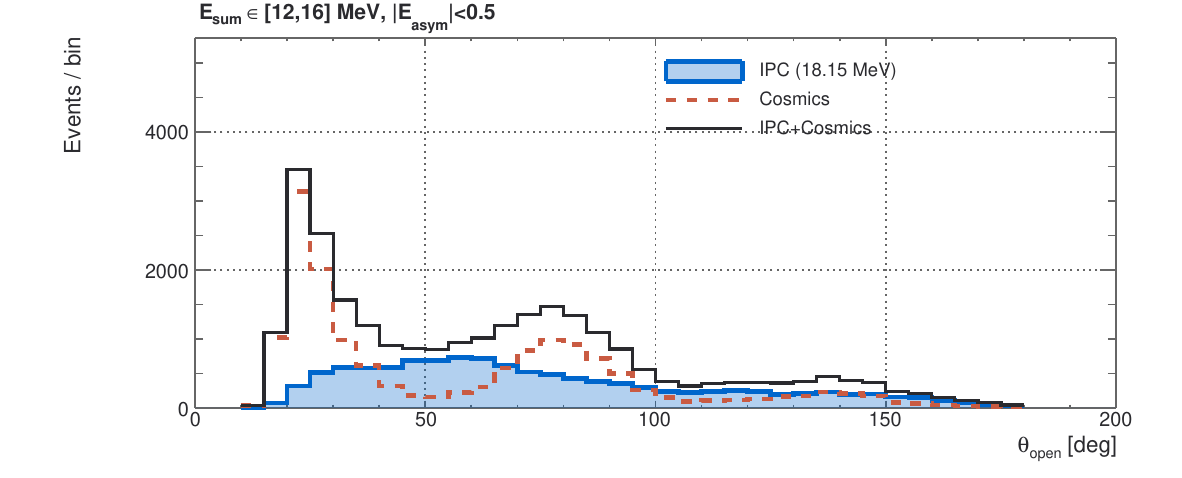}
  \caption{\label{fig:signal_region}
    $\thetaopen$ distributions in the signal window
    $\Esum \in [16, 20]$~MeV (top) and in the background window $\Esum \in [12, 16]$~MeV (bottom) with $|\Easym| < 0.5$) for IPC (solid blue filled),
    cosmics (dashed red), and their sum (solid black), normalized to 300~hours.
    In [16, 20]~MeV, the angular excess near $140^\circ$ in the sum originates in the
    cosmic component and its amplitude is approximately equal to the IPC one
    in this window. In [12, 16]~MeV, this cosmic excess has mostly disappeared.}
\end{figure}

Figure~\ref{fig:signal_region}~(top) shows the $\thetaopen$ distributions in
the signal window ($\Esum \in [16, 20]$~MeV, $|\Easym| < 0.5$) for the IPC and cosmic components separately, and their normalized sum. The IPC distribution, while being shaped by the spectrometer acceptance, is monotonically decreasing from $50^\circ$ onwards. The cosmic component carries the angular concentration near $120^\circ-160^\circ$. The two components contribute at roughly equal rates in this angular region. In the combined distribution, this produces a wide bump-like structure peaking at 140$^\circ$.
Again, it should be noted that this energy window also selects cosmic rays hitting $\Delta\phi = 60^\circ$ and $\Delta\phi = 90^\circ$ adjacent arms, leading to opening angles near $20^\circ$ and $60^\circ$. In this low-angle region, however, the larger IPC and EPC rates dilute the relative weight of cosmic events relative to the $140^\circ$ region.

\subsection{\label{sec:asymmetry_5arm}Energy Asymmetry and Background Cross-Checks}

For $|\Easym| > 0.5$ the cosmic excess near $140^\circ$ is slightly reduced relative to the symmetric sample, because cosmic tracks that produce strongly unequal
scintillator energies are geometrically less likely to connect the $\Delta\phi = 150^\circ$ arm pair.

In the background window $\Esum \in [12, 16]$~MeV, shown in Figure~\ref{fig:signal_region}~(bottom), the cosmic component near $140^\circ$ is largely reduced, consistent with the background region used by Atomki as an internal cross-check. It should be noted that the $^8$Be(15.1~MeV) IPC from the transition to the first excited state is not simulated here but would further decrease the cosmic-to-IPC ratio in the [12, 16]~MeV window.

% ─────────────────────────────────────────────────────────────────────────────
\section{\label{sec:discussion}Discussion}

\subsection{\label{sec:caveats}Limitations}

The present study does not include beam-related backgrounds (Compton
electrons, neutron-induced recoils) which may also contribute to the
two-arm coincidence rate. The cosmic
flux normalization carries an uncertainty of order 20--30\% from the
Gaisser parametrization and the neglect of geomagnetic cutoff effects
at the Debrecen latitude;
this affects the absolute number of events but not the shape of the
$\thetaopen$ distribution, which is what drives the angular structure
discussed above. The simulation also omits the beam pipe and target assembly; its effect on the cosmic background shape is expected to be small given the thinness of the material. 

This work also raises the question of how these considerations bear on other
searches for the $X17$ anomaly. PADME and
Abraamyan~et al. report positive signals despite being immune
to the cosmic-ray background presented here, favoring a non-cosmic
interpretation of the $X17$ anomaly. MEG~II, whose magnetic
spectrometer is likewise unaffected by this background, instead
reports a null result. These results are thus in tension, leaving the
anomaly's origin unresolved. The result of Anh~et al. is
ambiguous: an IPC experiment whose apparatus, unlike the others, could
be exposed to cosmics, with the degree of exposure highly
geometry-dependent. A dedicated simulation of cosmic-ray background in
that apparatus, along the lines presented here, would be particularly
valuable in resolving the anomaly's origin.

A further limitation concerns two other ATOMKI measurements using related apparatus.
The $^{12}$C measurement~\cite{Krasznahorkay2022c12} used the same six-arm geometry
simulated in Sec.~\ref{sec:6arm} but reported excesses extending above $\thetaopen\simeq150^\circ$
at $\Esum\simeq20$~MeV, a region our cosmic simulation does not reproduce.
The Giant Dipole Resonance measurement~\cite{Krasznahorkay2023gdr} instead used a two-telescope apparatus with a $110^\circ$ spacing, which should suppress two-arm cosmic coincidences relative to the five- and six-arm spectrometers. A dedicated simulation is left for future work.

Finally,  the beam-energy dependence shown in Fig.~2 of Ref.~\cite{Krasznahorkay2016} is not consistent with
a purely cosmic origin: the $^8$Be excess in the
$\thetaopen$ distribution appears and disappears as the proton beam energy
is tuned on and off the 1030~keV resonance. This
observation constrains any cosmic-background explanation.

\subsection{\label{sec:subtraction}Sensitivity to Cosmic Subtraction}

%The Atomki group estimates the cosmic background by recording beam-off data and normalising it to the beam-on sample in the region $\Esum > 25$~MeV, where no IPC signal is expected. The normalised beam-off $\thetaopen$ histogram is then subtracted from the beam-on histogram at the signal energy to isolate the physics signal. If the scintillator gain drifts with time or just differs between beam-on and beam-off conditions --- as can occur due to rate-induced effects --- the reconstructed energy scale shifts between the two datasets, so that the beam-off $\thetaopen$ distribution at $\Esum \simeq 18$~MeV no longer represents the same population of cosmic tracks as the beam-on sample at the same nominal energy. Because of the strong correlation between $\Esum$ and $\thetaopen$ demonstrated in Fig.~\ref{fig:esum_angle_2d}, such an energy scale difference translates directly into a shape mismatch in the subtracted angular distribution, potentially leaving a spurious residual.
The Atomki group estimates the cosmic background by recording beam-off data and normalising it to the beam-on sample in the region $\Esum > 25$~MeV~\cite{Krasznahorkay2018}, where no IPC signal is expected. The normalised beam-off $\thetaopen$ histogram is then subtracted from the beam-on histogram at the signal energy to isolate the physics signal. Among other possible effects, gain drifts are prone to producing differences between beam-on and beam-off data-taking. Given the strong correlation between $\Esum$ and $\thetaopen$ demonstrated in Fig.~\ref{fig:esum_angle_2d}, it could translate into a shape mismatch in the subtracted angular distribution, potentially leaving a spurious residual.
The normalization study in Sec.~\ref{sec:normalization}
shows that cosmic events represent more than 60\% of all two-arm coincidences in the $^8$Be signal window ($\Esum \in [16, 20]$~MeV and $\thetaopen$ in  $[120^\circ-160^\circ]$). Under these conditions, a relatively small error on the cosmic normalization could generate a residual angular structure of comparable amplitude to the reported excess. Quantifying the achievable calibration stability between beam-on and beam-off periods, and characterising possible gain drifts of the plastic scintillators photosensors under varying beam conditions, are therefore critical for this measurement. This reasoning assumes that the Atomki trigger or offline analysis does not anti-select cosmic-ray events through coincidence requirements, which would suppress their contamination to an accidental background level.

\subsection{\label{sec:geometry}Geometry-Driven Angular Structure}

The locations of the angular concentrations found in the simulation are
determined by the detector arm spacing and the E-plastics sizes. For the six-arm
layout with uniform $60^\circ$ steps, the next-to-adjacent pair geometry
($\Delta\phi = 120^\circ$) subtends $\thetaopen \simeq 140^\circ$ in the
transverse plane; the scintillator energy range then selects which muon
trajectories contribute, shifting the peak between $120^\circ$ and
$140^\circ$ as a function of $\Esum$. Any discrete multi-arm spectrometer
with azimuthal spacings of this order, lacking an active cosmic veto,
will produce qualitatively similar structures in the $(\thetaopen, \Esum)$
plane. Some variation in the exact radial positions of the simulated
detectors with respect to the true apparatus is possible, and would slightly shift
the precise peak locations and their $\Esum$ dependence; however,
this does not affect the central point of this work, namely that some
degree of $\thetaopen$--$\Esum$ correlation from cosmic coincidences is an unavoidable consequence of this class of discrete, multi-arm geometry and that these coincidences can mimic the topology of a light boson decay. %Ref.~\cite{Krasznahorkay2018} shows some measured cosmic spectra with 5-arm and 6-arm geometry. While the analysis details are not given, preventing a direct comparison with this work

\subsection{\label{sec:cross_section}Cross-sections}
An observation worth noting concerns the pattern of excesses across the different nuclear reactions studied at Atomki. Measurements with high cross-section transitions, such as the $^8$Be(17.6) ground-state transition~\cite{Krasznahorkay2016} and the $^{16}$O reaction produced by fluorine~\cite{Krasznahorkay2016} in the target, show no anomalous excess above the IPC expectation. By contrast, the reported excesses appear in lower cross-section measurements: the $^8$Be(18.1) transition and the $^4$He reaction. This pattern is consistent with a dependence on the signal-to-cosmic ratio: in high cross-section measurements, the IPC rate is large relative to the cosmic coincidence rate, and the impact of any imprecision in the cosmic subtraction on the final angular distribution is correspondingly small. In the lower cross-section cases, the cosmic contribution represents a larger fraction of the total two-arm event rate in the signal window --- as quantified in Sec.~\ref{sec:normalization} for the five-arm setup --- and the result becomes more sensitive to the cosmic background characterisation and subtraction.

% ─────────────────────────────────────────────────────────────────────────────
\section{\label{sec:conclusions}Conclusions}

We have simulated cosmic-ray muon backgrounds in the 2016 five-arm and
the 2021 six-arm Atomki pair spectrometers using \geant{} 11.1 with
full detector geometry and the Gaisser muon spectrum. The main findings
are as follows.

The discrete azimuthal arrangement of detector arms in both setups
produces preferred $\thetaopen$ values for two-arm cosmic coincidences.
In the six-arm geometry, the next-to-adjacent pair topology ($\Delta\phi = 120^\circ$) produces a peak near $140^\circ$ at the $^8$Be transition energy and near $120^\circ$ at the $^4$He transition energy, illustrating the strong energy deposit dependence of the cosmic-ray background. These peak positions are consistent with those reported by Atomki in Refs.~\cite{Krasznahorkay2016,Krasznahorkay2021}. Furthermore, the suppression of these structures for energy-asymmetric pairs arises naturally from the muon trajectory and matches the behavior of the reported excesses. Finally, the peaks disappear when the $\Esum$ window is shifted below the respective transition energies of $^8$Be and $^4$He. Together, these three characteristic features of the cosmic-ray background can mimic the topology of the excesses reported by Atomki.

For the five-arm setup, after normalizing $^8$Be IPC and cosmic backgrounds to $\sim$300~hours of data-taking at $I_p = 1~\mu$A, the cosmic contribution is found to be dominant in the $^8$Be signal window. The excess is reduced at $\Esum$ below the $^8$Be transition energy. It places a strong
accuracy requirement on Atomki's cosmic subtraction procedure. A similar background-control challenge is expected for the $^4$He measurement.

These findings do not by themselves settle the question of the origin of the Atomki excesses, as the simulation does not reproduce the actual beam-off data against which the Atomki group normalizes the cosmic contribution. They do, however, demonstrate that the cosmic muon background in these spectrometers has angular and energy features
that directly overlap with the signal selection, and that the fraction of cosmic events in the signal window is large enough that errors in the normalization procedure would produce residuals of the reported magnitude. Future measurements ---whether at
Atomki or elsewhere---would benefit significantly from an active cosmic
veto, extended beam-off statistics collected at the same energy as the
physics data, and a published characterisation of the
$(\thetaopen, \Esum)$ distribution in the beam-off sample.

% ─────────────────────────────────────────────────────────────────────────────
\begin{acknowledgments}
The authors thank Marco Mancini, Benito Góngora-Servín and colleagues at INFN Pisa for useful discussions. Simulations used \geant{} 11.1 and
ROOT~6. Computing resources were provided by the Paul Scherrer Institute.
\end{acknowledgments}

% ─────────────────────────────────────────────────────────────────────────────
%\appendix

%\section{\label{app:geometry}Geometry Parameters}

% ─────────────────────────────────────────────────────────────────────────────
\bibliography{references}

@article{Krasznahorkay2016,
  author  = {Krasznahork{\'a}y, A. J. and others},
  title   = {Observation of Anomalous Internal Pair Creation in
             ${}^8\mathrm{Be}$: A Possible Indication of a Light,
             Neutral Boson},
  journal = {Phys. Rev. Lett.},
  volume  = {116},
  pages   = {042501},
  year    = {2016},
  doi     = {10.1103/PhysRevLett.116.042501}
}

@article{Krasznahorkay2018,
doi = {10.1088/1742-6596/1056/1/012028},
url = {https://dx.doi.org/10.1088/1742-6596/1056/1/012028},
year = {2018},
month = {Jul},
publisher = {IOP Publishing},
volume = {1056},
number = {1},
pages = {012028},
author = {Krasznahork{\'a}y, A. J. and others},
title = {{New results on the $^8$Be anomaly}},
journal = {Journal of Physics: Conference Series}
}

@article{Krasznahorkay2021,
  author  = {Krasznahork{\'a}y, A. J. and others},
  title   = {{New anomaly observed in ${}^4\mathrm{He}$ supports the
             existence of the X17 particle}},
  journal = {Phys. Rev. C},
  volume  = {104},
  pages   = {044003},
  year    = {2021},
  doi     = {10.1103/PhysRevC.104.044003}
}

@article{Krasznahorkay2022c12,
  title = {{New anomaly observed in $^{12}\mathrm{C}$ supports the existence and the vector character of the hypothetical X17 boson}},
  author = {Krasznahork{\'a}y, A. J. and others},
  journal = {Phys. Rev. C},
  volume = {106},
  issue = {6},
  pages = {L061601},
  numpages = {5},
  year = {2022},
  month = {Dec},
  publisher = {American Physical Society},
  doi = {10.1103/PhysRevC.106.L061601},
  url = {https://link.aps.org/doi/10.1103/PhysRevC.106.L061601}
}

@misc{Krasznahorkay2023gdr,
      title={{Observation of the X17 anomaly in the decay of the Giant Dipole Resonance of $^8$Be}}, 
      author={Krasznahork{\'a}y, A. J. and others},
      year={2023},
      eprint={2308.06473},
      archivePrefix={arXiv},
      primaryClass={nucl-ex},
      url={https://arxiv.org/abs/2308.06473}, 
}

@article{Gulyas2016,
  author  = {Guly{\'a}s, J. and others},
  title   = {A pair spectrometer for measuring multipolarities of
             energetic nuclear transitions},
  journal = {Nucl. Instrum. Meth. A},
  volume  = {808},
  pages   = {21--28},
  year    = {2016},
  doi     = {10.1016/j.nima.2015.11.009}
}

@article{Feng2016,
  author  = {Feng, J. L. and others},
  title   = {Protophobic Fifth-Force Interpretation of the Observed
             Anomaly in ${}^8\mathrm{Be}$ Nuclear Transitions},
  journal = {Phys. Rev. Lett.},
  volume  = {117},
  pages   = {071803},
  year    = {2016},
  doi     = {10.1103/PhysRevLett.117.071803}
}

@article{Feng2017,
  title = {{Particle physics models for the 17 MeV anomaly in beryllium nuclear decays}},
  author = {Feng, Jonathan L. and others},
  journal = {Physical Review D},
  volume = {95},
  issue = {3},
  pages = {035017},
  numpages = {25},
  year = {2017},
  month = {Feb},
  publisher = {American Physical Society},
  doi = {10.1103/PhysRevD.95.035017},
  url = {https://link.aps.org/doi/10.1103/PhysRevD.95.035017}
}

@article{Feng2020,
  author  = {Feng, J. L. and others},
  title   = {Particle physics models for the $17~\mathrm{MeV}$ anomaly
             in beryllium nuclear decays},
  journal = {Phys. Rev. D},
  volume  = {102},
  pages   = {036016},
  year    = {2020},
  doi     = {10.1103/PhysRevD.102.036016}
}

@article{Ellwanger2016,
author={Ellwanger, Ulrich and Moretti, Stefano},
title={{Possible explanation of the electron positron anomaly at 17 MeV in $^8$Be transitions through a light pseudoscalar}},
journal={Journal of High Energy Physics},
year={2016},
month={Nov},
day={08},
volume={2016},
number={11},
pages={39},
issn={1029-8479},
doi={10.1007/JHEP11(2016)039},
url={https://doi.org/10.1007/JHEP11(2016)039}
}

@article{Viviani2022,
  title = {{$X17$ boson and the $^{3}\mathrm{H}(\mathrm{p},\mathrm{e}^{+}\mathrm{e}^{\ensuremath{-}})^{4}\mathrm{He}$ and $^{3}\mathrm{He}(\mathrm{n},\mathrm{e}^{+}\mathrm{e}^{\ensuremath{-}})^{4}\mathrm{He}$ processes: A theoretical analysis}},
  author = {Viviani, M. and others},
  journal = {Physical Review C},
  volume = {105},
  issue = {1},
  pages = {014001},
  numpages = {30},
  year = {2022},
  month = {Jan},
  publisher = {American Physical Society},
  doi = {10.1103/PhysRevC.105.014001},
  url = {https://link.aps.org/doi/10.1103/PhysRevC.105.014001}
}

@misc{Aleksejevs2021,
      title={{A Standard Model Explanation for the "ATOMKI Anomaly"}}, 
      author={A. Aleksejevs and S. Barkanova and Yu. G. Kolomensky and B. Sheff},
      year={2021},
      eprint={2102.01127},
      archivePrefix={arXiv},
      primaryClass={hep-ph},
    note={arXiv:2102.01127}

}

@article{Koch2021,
title = {{X17: A new force, or evidence for a hard $\gamma+\gamma$ process?}},
journal = {Nuclear Physics A},
volume = {1008},
pages = {122143},
year = {2021},
issn = {0375-9474},
doi = {https://doi.org/10.1016/j.nuclphysa.2021.122143},
url = {https://www.sciencedirect.com/science/article/pii/S0375947421000087},
author = {Benjamin Koch}
}

@article{Afanaciev2025,
author={{K. Afanaciev \textit{et al.} (MEG-II Collaboration)}},
  title   = {{Search for the X17 particle in $^{7}\mathrm{Li}(p,e^{+}e^{-})^{8}\mathrm{Be}$ processes with the MEG II detector}},
  journal = {The European Physical Journal C},
  year    = {2025},
  volume  = {85},
  number  = {7},
  pages   = {763},
  doi     = {10.1140/epjc/s10052-025-14345-0},
  url     = {https://doi.org/10.1140/epjc/s10052-025-14345-0}
}

@Article{Afanaciev2024,
author={{K. Afanaciev \textit{et al.} (MEG-II Collaboration)}},
title={{Operation and performance of the MEG II detector}},
journal={The European Physical Journal C},
year={2024},
month={Feb},
day={26},
volume={84},
number={2},
pages={190},
issn={1434-6052},
doi={10.1140/epjc/s10052-024-12415-3},
url={https://doi.org/10.1140/epjc/s10052-024-12415-3}
}

@article{CDCHpaper,
author={Baldini, A. M.
and others},
title={{Performances of a new generation tracking detector: the MEG II cylindrical drift chamber}},
journal={The European Physical Journal C},
year={2024},
month={May},
day={07},
volume={84},
number={5},
pages={473},
issn={1434-6052},
doi={10.1140/epjc/s10052-024-12711-y},
url={https://doi.org/10.1140/epjc/s10052-024-12711-y}
}

@Article{Abraamyan2023,
author={Abraamyan, Kh. U.
and others},
title={{Observation of Structures at $\sim$17 and $\sim$38~MeV/c$^2$ in the
$\gamma\gamma$ Invariant Mass Spectrum in dCu Collisions at a Momentum of
3.8~GeV/c per Nucleon}},
journal={Physics of Particles and Nuclei},
year={2024},
month={Aug},
day={01},
volume={55},
number={4},
pages={868-873},
issn={1531-8559},
doi={10.1134/S1063779624700412},
url={https://doi.org/10.1134/S1063779624700412}
}

@article{Anh2024,
AUTHOR = {Anh, Tran The and others},
title = {{Checking the $^8$Be Anomaly with a Two-Arm Electron Positron Pair Spectrometer}},
JOURNAL = {Universe},
VOLUME = {10},
YEAR = {2024},
NUMBER = {4},
ARTICLE-NUMBER = {168},
URL = {https://www.mdpi.com/2218-1997/10/4/168},
ISSN = {2218-1997},
DOI = {10.3390/universe10040168}
}

@article{Batley2015,
title = {{Search for the dark photon in $\pi^0$ decays}},
journal = {Physics Letters B},
volume = {746},
pages = {178-185},
year = {2015},
issn = {0370-2693},
doi = {https://doi.org/10.1016/j.physletb.2015.04.068},
url = {https://www.sciencedirect.com/science/article/pii/S0370269315003342},
author = {{J. R. Batley \textit{et al.} (NA48/2 Collaboration)}}
}

@article{Banerjee2020,
  title = {{Improved limits on a hypothetical $X(16.7)$ boson and a dark photon decaying into $\mathrm{e}^{+}\mathrm{e}^{\ensuremath{-}}$ pairs}},
  author = {{D. Banerjee \textit{et al.} (NA64 Collaboration)}},
  collaboration = {The NA64 Collaboration},
  journal = {Physical Review D},
  volume = {101},
  issue = {7},
  pages = {071101},
  numpages = {7},
  year = {2020},
  month = {Apr},
  publisher = {American Physical Society},
  doi = {10.1103/PhysRevD.101.071101},
  url = {https://link.aps.org/doi/10.1103/PhysRevD.101.071101}
}

@Article{Padme2025,
author  = {{F. Bossi \textit{et al.} (PADME Collaboration)}},
title={{Search for a new 17 MeV resonance via e$^+$e$^-$ annihilation with the PADME experiment}},
journal={Journal of High Energy Physics},
year={2025},
month={Nov},
day={04},
volume={2025},
number={11},
pages={7},
issn={1029-8479},
doi={10.1007/JHEP11(2025)007},
url={https://doi.org/10.1007/JHEP11(2025)007}
}

@article{Geant4_2003,
  author  = {Agostinelli, S. and others},
  title   = {{Geant4} --- a simulation toolkit},
  journal = {Nucl. Instrum. Meth. A},
  volume  = {506},
  pages   = {250--303},
  year    = {2003},
  doi     = {10.1016/S0168-9002(03)01368-8}
}

@article{Geant4_2006,
  author  = {Allison, J. and others},
  title   = {{Geant4} developments and applications},
  journal = {IEEE Trans. Nucl. Sci.},
  volume  = {53},
  pages   = {270--278},
  year    = {2006},
  doi     = {10.1109/TNS.2006.869826}
}

@article{Geant4_2016,
  author  = {Allison, J. and others},
  title   = {Recent developments in {Geant4}},
  journal = {Nucl. Instrum. Meth. A},
  volume  = {835},
  pages   = {186--225},
  year    = {2016},
  doi     = {10.1016/j.nima.2016.06.125}
}

@article{Zhang2017,
title = {{Can nuclear physics explain the anomaly observed in the internal pair production in the Beryllium-8 nucleus?}},
journal = {Physics Letters B},
volume = {773},
pages = {159-165},
year = {2017},
issn = {0370-2693},
doi = {https://doi.org/10.1016/j.physletb.2017.08.013},
url = {https://www.sciencedirect.com/science/article/pii/S0370269317306342},
author = {Xilin Zhang and Gerald A. Miller}
}

@article{Gysbers2023,
  title = {{$\mathit{Ab~initio}$ investigation of the $^{7}\mathrm{Li}(\mathrm{p},\mathrm{e}^{+}\mathrm{e}^{\ensuremath{-}})^{8}\mathrm{Be}$ process and the X17 boson}},
  author = {Gysbers, P. and Navr\'atil, P. and Kravvaris, K. and Hupin, G. and Quaglioni, S.},
  journal = {Physical Review C},
  volume = {110},
  issue = {1},
  pages = {015503},
  numpages = {25},
  year = {2024},
  month = {Jul},
  publisher = {American Physical Society},
  doi = {10.1103/PhysRevC.110.015503},
  url = {https://link.aps.org/doi/10.1103/PhysRevC.110.015503}
}

@article{Rose1949,
  title = {{Internal Pair Formation}},
  author = {Rose, M. E.},
  journal = {Physical Review},
  volume = {76},
  issue = {5},
  pages = {678--681},
  numpages = {0},
  year = {1949},
  month = {Sep},
  publisher = {American Physical Society},
  doi = {10.1103/PhysRev.76.678},
  url = {https://link.aps.org/doi/10.1103/PhysRev.76.678}
}

@article{Rose1951,
  title = {{The Internal Conversion Coefficients. I: The $K$-Shell}},
  author = {Rose, M. E. and Goertzel, G. H. and Spinrad, B. I. and Harr, J. and Strong, P.},
  journal = {Physical Review},
  volume = {83},
  issue = {1},
  pages = {79--87},
  numpages = {0},
  year = {1951},
  month = {Jul},
  publisher = {American Physical Society},
  doi = {10.1103/PhysRev.83.79},
  url = {https://link.aps.org/doi/10.1103/PhysRev.83.79}
}

@incollection{Rose1966,
title = {{Chapter II - Internal Conversion Theory}},
editor = {Joseph H. Hamilton},
booktitle = {{Internal Conversion Processes}},
publisher = {Academic Press},
pages = {15-33},
year = {1966},
isbn = {978-0-12-395610-1},
doi = {https://doi.org/10.1016/B978-0-12-395610-1.50006-7},
url = {https://www.sciencedirect.com/science/article/pii/B9780123956101500067},
author = {Rose, M. E.},
}

@book{Gaisser,
  author    = {Gaisser, T. K.},
  title     = {Cosmic Rays and Particle Physics},
  publisher = {Cambridge University Press},
  year      = {1990},
  isbn      = {978-0521339316}
}

@article{ROOT,
  author  = {Brun, R. and Rademakers, F.},
  title   = {{ROOT} --- An object oriented data analysis framework},
  journal = {Nucl. Instrum. Meth. A},
  volume  = {389},
  pages   = {81--86},
  year    = {1997},
  doi     = {10.1016/S0168-9002(97)00048-X}
}

\end{document}